# Confirmation of general relativity on large scales from weak lensing and galaxy velocities[1]

Reinabelle Reyes[1], Rachel Mandelbaum[1], Uros Seljak[2-4], Tobias Baldauf[2], James E. Gunn[1], Lucas Lombriser[2], Robert E. Smith[2]

[1]*Princeton University Observatory, Peyton Hall, Princeton, NJ 08544 USA*

[2]*Institute for Theoretical Physics, University of Zurich, Zurich, 8057, Switzerland*

[3]*Physics and Astronomy Department and Lawrence Berkeley National Laboratory, University of California-Berkeley, CA 94720 USA*

[4]*Institute for Early Universe, Ewha University, Seoul, S. Korea*

**Einstein's general relativity (GR) is the theory of gravity underpinning our understanding of the Universe, encapsulated in the standard cosmological model (ΛCDM). To explain observations showing that the Universe is undergoing accelerated expansion[1,2], ΛCDM posits the existence of a gravitationally repulsive fluid, called dark energy (in addition to ordinary matter and dark matter). Alternatively, the breakdown of GR on cosmological length scales could also explain the cosmic acceleration. Indeed, modifications to GR have been proposed as alternatives to dark energy[3,4], as well as to dark matter.[5,6] These modified gravity theories are designed to explain the observed expansion history, so the only way to test them is to study cosmological perturbations (deviations of the matter density from its mean value). This is a non-trivial task, compounded by our lack of *a priori* knowledge of relevant astrophysical parameters.[7,8] Here, we successfully measure the probe of gravity[9] $E_G$ that is robust to these uncertainties. Under GR+ΛCDM, $E_G$ should approximately equal 0.4. We find $E_G$ = 0.39±0.06 at**

---

**redshift 0.3, thus confirming the prediction of GR on length scales of tens of megaparsecs (~$10^{23}$ meters). Moreover, this result establishes a new, model-independent test of any modified gravity theory.**

The probe of gravity $E_G$ combines three different probes of large-scale structure: galaxy-galaxy lensing, galaxy clustering, and galaxy velocities derived from galaxy clustering in redshift space. Galaxy-galaxy lensing arises from the gravitational deflection of light and is sensitive to the sum of the two scalar potentials in the gravitational metric, $\psi$ and $\varphi$[10] (the metric quantifies the distance between any two points in space-time). On the other hand, galaxy clustering arises from the gravitational attraction of matter and is sensitive only to the Newtonian potential $\varphi$. Thus, galaxy-galaxy lensing and galaxy clustering together probe the relationship between the two potentials. In the GR+ΛCDM model, the two are equal at late times in the absence of anisotropic stress, but in modified gravity theories, there can be a systematic difference between the two, known as "gravitational slip",[11] even in the absence of anisotropic stress. The third probe, the galaxy velocity field, is sensitive to the rate of growth of structure. Structure growth induces coherent streaming motions, which lead to anisotropy in the clustering pattern of galaxies in redshift space, also known as galaxy redshift distortions[12]. The rate of growth of structure, in turn, depends on the theory of gravity, and in general, modified gravity theories will have a different structure growth relative to GR[13]. Thus, $E_G$ is sensitive to modifications to gravity that manifest in differences in the rate of growth of structure, in the gravitational slip, or in both.

For a model-independent constraint, it is important that we obtain $E_G$ by applying these different probes to the same set of galaxies. Individually, galaxy-galaxy lensing, galaxy clustering and galaxy redshift distortions are strongly sensitive to the galaxy bias $b$, which connects galaxy density perturbations to the underlying matter density perturbations. On large enough scales, the galaxy and matter perturbations are roughly linearly related by the galaxy bias[14], but the value of the bias itself is poorly constrained. Moreover, galaxy-galaxy lensing and galaxy clustering depend on the amplitude of the matter perturbations *A*, which we also do not know *a priori*. However, the combination of these quantities in $E_G$ is such that both nuisance parameters cancel out. Thus, unlike in previous analyses[15], we do not require additional observations and assumptions to estimate the galaxy bias, and are able to obtain more robust results.

We use a sample of 70,205 luminous red galaxies[16] (LRGs) from the Sloan Digital Sky Survey (SDSS)[17], a homogeneous dataset ideal for the study of large-scale structure. The galaxies have been selected according to the same criteria as in Eisenstein *et al.*[18] They cover an area of 5215 sq. degrees and a range of redshifts $z = 0.16 - 0.47$. The redshift $z = \lambda_{meas}/\lambda_{emis} - 1$ of the radiation emitted by a distant galaxy is a measure of the time of emission. The redshift of our galaxy sample, $z = 0.32$, corresponds to a lookback time of 3.5 billion years, when the universe was about 77 per cent of its current size, and is already well into the accelerated phase of the cosmic expansion. The sample also spans a large comoving volume, $1.02h^{-3}$ Gpc$^3$, where the Hubble constant $H_0 = 100h$ km s$^{-1}$ Mpc$^{-1}$, and 1 Gpc (giga-parsec) = 1000 Mpc (mega-parsec) = $3.086 \times 10^{25}$ m.

Tegmark *et al.*[19] measured the anisotropy in the power spectra of an equally selected sample of LRGs to determine the redshift distortion parameter $\beta \equiv f(z)/b$, where $f(z)$ is the logarithmic linear growth rate of structure at redshift *z*. Their analysis found $\beta = 0.309 \pm 0.035$ on large scales and at $z = 0.32$. In this work, we use this result for $\beta$, together with new measurements of the galaxy-galaxy lensing and galaxy clustering signals of the full LRG sample, to determine $E_G$ at mega-parsec scales and effective redshift of $z = 0.32$.

Galaxy-galaxy lensing is the slight distortion of shapes of "source" galaxies in the background of "lens" galaxies due to the gravitational deflection of light by matter along the line of sight. Using shape measurements[20] of more than 30 million source galaxies covering the same area as the LRG sample, we calculate the lensing signal around the LRGs and stack them together to achieve a high signal-to-noise ratio (see Supplementary Information for details). In the weak lensing limit, we denote the lensing signal profile $\Delta\Sigma_{gm}(R)$ since it is directly proportional to the projected surface mass density contrast, $\overline{\Sigma}_{gm}(<R) - \Sigma_{gm}(R)$, where $\Sigma_{gm}(R)$ is the projected surface mass density at $R$, and $\overline{\Sigma}_{gm}(<R)$ is its mean value within $R$. Figure 1a shows the average galaxy-galaxy lensing profile measured from the LRG sample for scales $R = 1.5 - 47 h^{-1}$ Mpc.

On the scales we consider, galaxy perturbations are not perfectly proportional to the matter density perturbations[21], so we seek to minimize the effect of stochasticity and scale dependence of galaxy bias on our measurement of $E_G$. To do so, we introduce a new statistic $\Upsilon_{gm}(R)$, the galaxy-matter annular differential surface density,

$$\Upsilon_{gm}(R) \equiv \Delta\Sigma_{gm}(R) - \left(\frac{R_0}{R}\right)^2 \Delta\Sigma_{gm}(R_0) \tag{1a}$$

$$= \frac{2}{R^2}\int_{R_0}^{R} dR' R' \Sigma_{gm}(R') - \Sigma_{gm}(R) + \left(\frac{R_0}{R}\right)^2 \Sigma_{gm}(R_0). \tag{1b}$$

By construction, $\Upsilon_{gm}(R)$ does not include any contribution from length scales smaller than $R = R_0$. In practice, one must choose $R_0$ to be large enough to suppress the systematic effects arising from small scales, but small enough to preserve a high signal-to-noise ratio in the measurement.

The two-point galaxy correlation function is a basic measure of galaxy clustering due to gravitational attraction. We estimate it using the standard method of counting galaxy pairs and comparing the result with that for a randomly distributed sample[22]. We integrate this quantity along the line of sight to obtain the projected two-point galaxy correlation function $w_{gg}(R)$ (shown in Figure 1b). In parallel with equation (1b), we define $\Upsilon_{gg}(R)$, the galaxy-galaxy annular differential surface density, as

$$\Upsilon_{gg}(R) \equiv \rho_c \left[ \frac{2}{R^2} \int_{R_0}^{R} dR' R' w_{gg}(R') - w_{gg}(R) + \left(\frac{R_0}{R}\right)^2 w_{gg}(R_0) \right], \quad (2)$$

where $\rho_c = 3H^2/8\pi G$ is the critical matter density of the Universe. Both quantities, $\Upsilon_{gm}(R)$ and $\Upsilon_{gg}(R)$, have units of surface mass density.

We define the probe of gravity $E_G$ as a function of scale, as

$$E_G(R) = \frac{1}{\beta} \frac{\Upsilon_{gm}(R)}{\Upsilon_{gg}(R)}. \quad (3)$$

Both the galaxy bias and amplitude of fluctuations cancel out in this expression. In practice, $\Upsilon_{gm}(R)$ and $\Upsilon_{gg}(R)$ are calculated using equations (1a) and (2), respectively, with a minimum scale of $R_0 = 1.5 h^{-1}$ Mpc. We choose this minimum scale to be close to the typical virial radius of the halos of the most massive LRGs, above which we expect the galaxies to trace the dark matter, but our results are not very sensitive to this particular choice of $R_0$. To estimate errors in $E_G(R)$, while at the same time accounting for any correlations between radial bins, we use jackknife resampling of 34 galaxy subsamples from equal-area regions in the sky. To obtain numerical corrections to $E_G$ to account for the effect of scale-dependent bias and other systematic effects, we use a suite of dark matter simulations[23], which have been populated with galaxies using the halo occupation distribution (HOD) model[24] that best reproduces the observations (see Fig. 1 and Supplementary Information). The correction factors that we obtain are well below the statistical uncertainty in $E_G$.

Figure 2 shows our estimate of $E_G(R)$, with $1\sigma$ error bars that include the error in the measurement of $\beta$. Taking the average of $E_G(R)$ over $R = 10 - 50 h^{-1}$ Mpc, and accounting for correlations in the data, we find $\langle E_G \rangle = 0.392 \pm 0.065$ ($1\sigma$) (grey shaded region in Fig. 2). The 16 per cent error on $E_G$ is dominated by the 11 per cent statistical error on $\beta$ and the 12 per cent statistical error on the galaxy-galaxy lensing signal. In addition, we have about a 5 per cent lensing calibration uncertainty[20]. As detailed in the Supplementary Information, systematic effects on $E_G$ are least important on length scales $R > 10 h^{-1}$ Mpc, so the results are most robust there. We note that the average over $R = 2 - 50 h^{-1}$ Mpc yields a result consistent with that above, $\langle E_G \rangle = 0.40 \pm 0.07$.

The GR+$\Lambda$CDM model prediction is $E_G = \Omega_{m,0}/f(z) = 0.408 \pm 0.029$, where the matter density parameter today[25] is $\Omega_{m,0} = 0.2565 \pm 0.018$ and the logarithmic linear growth rate of structure $f(z) \approx \Omega_m(z)^{0.55} \approx 0.629$ at $z = 0.32$. The data are consistent with this prediction over the range of scales we consider (see solid line and labelled vertical bar in Fig. 2). Providing model independent constraints on gravitational slip is complicated, since gravitational slip will affect the rate of growth of structure. What is clear is that there is no evidence for a gravitational slip from our data. Thus, we find no deviation from GR on length scales $10^{11}$ times larger than those for which classical tests[26] have been performed.

We also compare our constraint on $E_G$ with predictions from two viable modified gravity theories—tensor-vector-scalar (TeVeS)[6] and $f(R)$[5] gravity (see labelled vertical bars in Fig. 2). $f(R)$ gravity models[27] that are designed to reproduce the observed cosmic expansion history with a specific model for gravitational slip predict a range of $E_G = 0.328 - 0.365$ (see Supplementary Information). The data favour slightly higher values, but are consistent with the predicted range. These models can be tested in the near future as limits on $E_G$ improve with larger data sets and better control of systematic errors from the next generation of galaxy surveys. Nevertheless, even with the current limits, we can tentatively rule out particular models. For example, the TeVeS model considered by Zhang *et al.*[9] predicts $E_G = 0.22$, lower than the observed value by more than $2.5\sigma$. Whether this result rules out the entire class of TeVeS models is an open issue[28] but in any case, it serves as a concrete demonstration that our measurement of $E_G$ presents a new and non-trivial challenge to both existing and future proposals of modifications to GR.

This work also serves as a proof of concept of the feasibility and power of this approach of probing gravity. We anticipate that our result will merely be the first among many determinations of $E_G$, each subsequently placing tighter limits and providing greater discriminatory power toward the goal of determining the nature of gravity, and concomitantly, the origin of the cosmic acceleration.

**Supplementary Information** accompanies the paper on **www.nature.com/nature**.

**Acknowledgements** R.M. was supported for the duration of this work by NASA through Hubble Fellowship grant #HST-HF-01199.02-A awarded by the Space Telescope Science Institute, which is operated by the Association of Universities for Research in Astronomy, Inc., for NASA, under contract NAS 5-26555. U.S. acknowledges the Swiss National Foundation under contract 200021-116696/1 and WCU grant R32-2008-000-10130-0. T.B. acknowledges support by a grant from the German National Academic Foundation during the initial phase of this project.

**Author Contributions** R.R., R.M., U.S. & J.E.G. worked on the observational analysis, with R.R. doing most of the computations. T.B. & R.E.S. worked on the numerical simulations, with T.B. calculating the correction factors used in this work. L.L. worked on the theoretical predictions for comparison with the observations.

**Author Information** Correspondence and requests for materials should be addressed to R.R. (e-mail: rreyes@astro.princeton.edu)

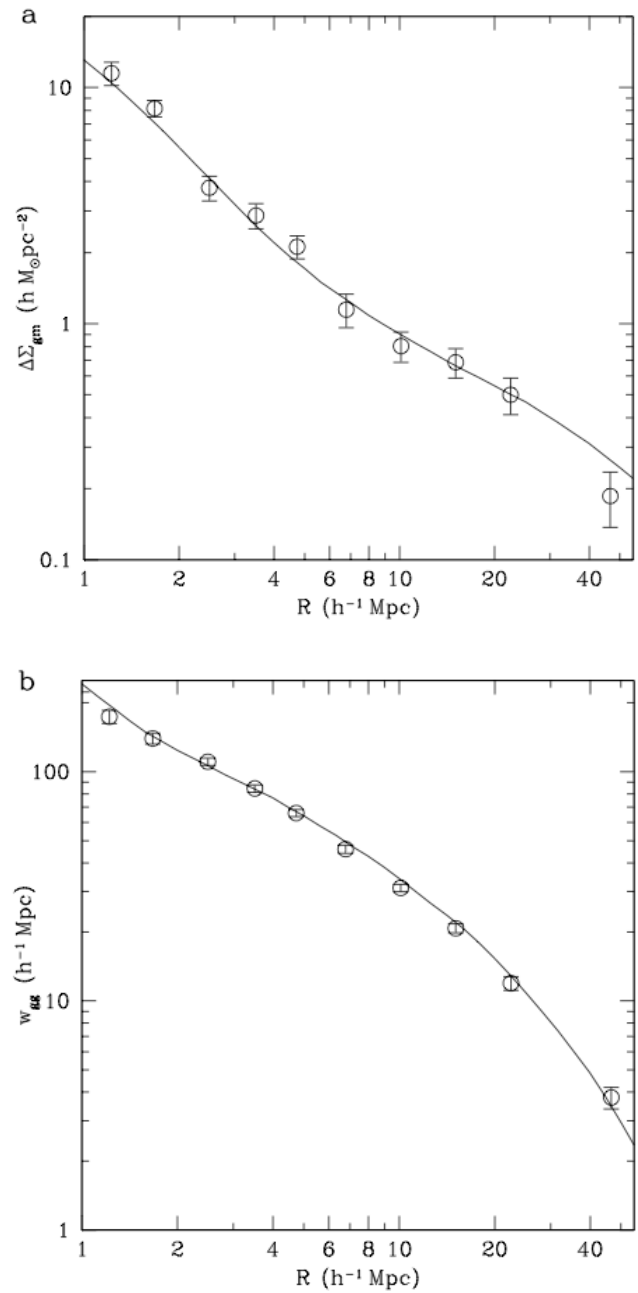

**Figure 1 | Probes of large-scale structure measured from ~70,000 luminous red galaxies (LRGs).** Observed radial profiles for two complementary probes, galaxy-galaxy lensing (**a**) and galaxy clustering (**b**) are shown for scales $R = 1.5 - 47 h^{-1}$ Mpc (open circles). The $1\sigma$ error bars (s.d.) are estimated from jackknife resampling of 34 equal-area regions in the sky. Profiles measured from mock galaxy catalogues are also shown (solid curves).

To generate the mock galaxy catalogues, we use a standard five-parameter halo occupation distribution (HOD) model— with two parameters related to the assignment of central galaxies, and three parameters related to the distribution of satellite galaxies (see Supplementary Information for more details). To fix the HOD model parameters, we require the galaxy number density to match the observed value, and find the best joint fit to the observed galaxy-galaxy lensing and galaxy clustering profiles. Despite this tuning, it is remarkable that this simple model is able to reproduce both the overall shape and particular features of the observed profiles.

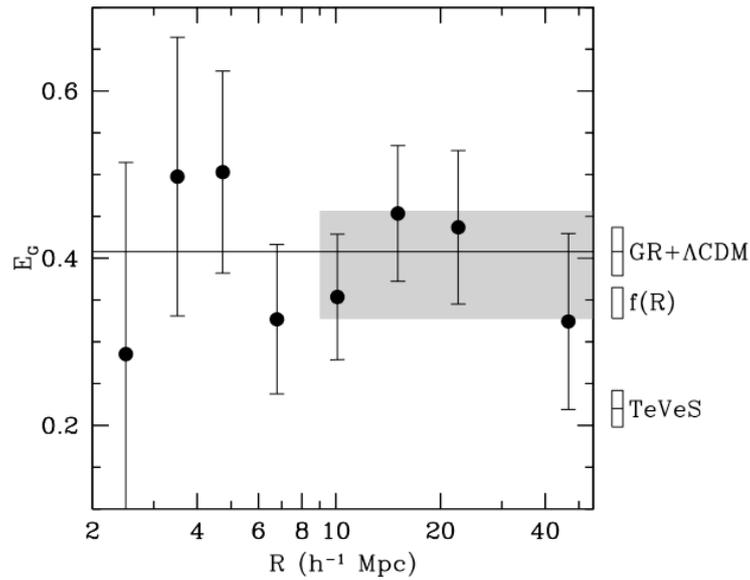

**Figure 2 | Comparison of observational constraints with predictions from GR and viable modified gravity theories.** Estimates of $E_G(R)$ are shown with 1σ error bars (s.d.) including the statistical error on the measurement[19] of $\beta$ (filled circles). The grey shaded region indicates the $1\sigma$ envelope of the mean $E_G$ over scales $R = 10 - 50 h^{-1}$ Mpc, where the systematic effects are least important (see Supplementary Information). The horizontal line shows the mean prediction of the GR+ΛCDM model, $E_G = \Omega_{m,0}/f$, for the effective redshift of the measurement, $z = 0.32$. On the right side of the panel, labelled vertical bars show the predicted ranges from three different gravity theories: (i) GR+ΛCDM ($E_G = 0.408 \pm 0.029\,(1\sigma)$), (ii) a class of cosmologically-interesting models in $f(R)$ theory with Compton wavelength parameters[27] $B_0 = 0.001 - 0.1$ ($E_G = 0.328 - 0.365$), and (iii) a TeVeS model[9] designed to match existing cosmological data and to produce a significant enhancement of the growth factor ($E_G = 0.22$, shown with a nominal error bar of 10 per cent for clarity).